\documentclass[aps,prl,twocolumn,showpacs,superscriptaddress]{revtex4}

\usepackage{amsfonts}
\usepackage{amsmath}
\usepackage{amssymb}
\usepackage{graphicx}
\usepackage{bm}

\begin{document}


\title{Intrinsic pinning on structural domains in underdoped single crystals of Ba(Fe$_{1-x}$Co$_x$)$_2$As$_2$}


\author{R.~Prozorov}
\email[Corresponding author: ]{prozorov@ameslab.gov}
\affiliation{Ames Laboratory, Ames, Iowa 50011, USA}
\affiliation{Department of Physics and Astronomy, Iowa State University, Ames, Iowa 50011, USA }

\author{M.~A.~Tanatar}
\affiliation{Ames Laboratory, Ames, Iowa 50011, USA}

\author{N.~Ni}

\affiliation{Ames Laboratory, Ames, Iowa 50011, USA}
\affiliation{Department of Physics and Astronomy, Iowa State University, Ames, Iowa 50011, USA }

\author{A.~Kreyssig}
\affiliation{Ames Laboratory, Ames, Iowa 50011, USA}

\author{S.~Nandi}
\affiliation{Ames Laboratory, Ames, Iowa 50011, USA}

\author{S.~L.~Bud'ko}
\affiliation{Ames Laboratory, Ames, Iowa 50011, USA}
\affiliation{Department of Physics and Astronomy, Iowa State University, Ames, Iowa 50011, USA }

\author{A.~I.~Goldman }
\affiliation{Ames Laboratory, Ames, Iowa 50011, USA}
\affiliation{Department of Physics and Astronomy, Iowa State University, Ames, Iowa 50011, USA }

\author{P.~C.~Canfield}
\affiliation{Ames Laboratory, Ames, Iowa 50011, USA}
\affiliation{Department of Physics and Astronomy, Iowa State University, Ames, Iowa 50011, USA }

\date{22 September 2009}


\begin{abstract}
Critical current density was studied in single crystals of Ba(Fe$_{1-x}$Co$_x$)$_2$As$_2$ for the values of $x$ spanning the entire doping phase diagram. A noticeable enhancement was found for slightly underdoped crystals with the peak at $x = 0.058$.  Using a combination of polarized-light imaging, x-ray diffraction and magnetic measurements we associate this behavior with the intrinsic pinning on structural domains in the orthorhombic phase. Domain walls extend throughout the sample thickness in the direction of vortices and act as extended pinning centers. With the increasing $x$ domain structure becomes more intertwined and fine due to a decrease of the orthorhombic distortion. This results in the energy landscape with maze-like spatial modulations favorable for pinning. This finding shows that iron-based pnictide superconductors, characterized by high values of the transition temperature, high upper critical fields, and low anisotropy may intrinsically have relatively high critical current densities.
\end{abstract}





\maketitle

\section{Introduction}

In type II superconductors magnetic field penetrates the bulk in the form of Abrikosov vortices. In the presence of electric current these vortices experience the Lorenz force, and, if not hindered, their motion leads to a non-zero resistance. Vortices can retain their positions due to pinning caused by local variations of the superconducting properties that lead to a position-dependent vortex energy. Therefore, electric current flows without dissipation, as long as its density, $j$, stays below some temperature and field dependent critical value, called critical current density, $j_c$ \cite{Bean}. The pinning strength is low in homogeneous high-quality single crystals, unless they have defects introduced either artificially or naturally \cite{Blatter}. There are different types of pinning centers ranging in size and dimensionality. Since vortices are linear objects threading the entire sample, they are more strongly pinned by extended defects. In highly anisotropic layered superconductors pinning is maximum when vortices lay parallel to the layers. However, the measured critical current density is determined by a much weaker pinning of vortices that are perpendicular to the superconducting layers \cite{Blatter}. The most efficient way to stop such vortices from bending and moving is to pin them by defects that extend perpendicular to the layers. Two-dimensional (planar) pinning centers can appear naturally, for example, as twin boundaries in YBa$_2$Cu$_3$O$_{7-y}$ (YBCO) \cite{vinnikov88} or in the orthorhombic/antiferromagnetic (AFM) phase in ErNi$_2$B$_2$C and HoNi$_2$B$_2$C \cite{Er-ortho,Er-decoration}. Linear pinning centers can be artificially created, for example, by heavy ion irradiation that produces columnar defects (see, e.g., Section IX in \cite{Blatter}). Whereas such defects lead to certain enhancement of pinning in high critical temperature cuprate superconductors (high-T$_c$ cuprates), their extreme anisotropy has been proven to be detrimental for technological use. The discovery of iron-based pnictide superconductors \cite{Hosono,Rotter} with relatively high transition temperatures, $T_c$, has inspired new hopes. Not only do these materials have very high upper critical fields comparable to cuprates \cite{highHc2}, they have relatively low anisotropy \cite{NiNiCo,Kano,low_anisotropy,singleton,tanatar_transport}, which is beneficial for applications. Nevertheless, the vortex properties of the iron pnictide superconductors were found to be similar to the cuprates and can be understood within the weak collective pinning approach \cite{Wen,Prozorov,Yamamoto,Kim,decoration,equalpenetration}. Not surprisingly, heavy ion irradiation of pnictides also leads to the enhancement of the critical current density \cite{Tamegai}.

In this work, we show that by fine tuning the composition in Ba(Fe$_{1-x}$Co$_x$)$_2$As$_2$ ("Co-Ba122") one can achieve a significant increase of pinning due to twin boundaries in the AFM/orthorhombic phase. Previously we directly imaged such boundaries in several AEFe$_2$As$_2$ (AE=Ca, Sr, Ba) parent compounds \cite{domains}. The effect of domain boundaries on the direction of vortex motion in Co-doped compositions was shown with SQUID microscopy \cite{Kirtley}. Here we demonstrate that these domains co-exist with superconductivity almost up to optimal doping and play the role of efficient extended pinning centers on the slightly underdoped side. Unlike YBCO, where the density of twin boundaries varies with sample and, in the cleanest samples, can be absent, in the pnictide single crystals a maze of fine robust domains always appears up to the optimal doping level and leads to a substantial, \textit{intrinsic pinning}.

\section{Experimental}

Single crystals of Ba(Fe$_{1-x}$Co$_x$)$_2$As$_2$ were grown from FeAs flux from a starting load of metallic Ba and FeAs and CoAs, as described in detail elsewhere \cite{NiNiCo}. Crystals were thick platelets with sizes as big as 12$\times$8$\times$1 mm$^3$ and large faces corresponding to the tetragonal (001) plane. Samples for magnetization and optical microscopy polarized-light imaging were cleaved with a razor blade into rectangular shaped platelets of typical dimensions, $(2-5)\times(2-5)\times(0.1-0.5)$ mm$^3$. 

To characterize the lattice chaqnge associated with the tetragonal-to-orthorhombic structural phase transition and to check the crystal perfection, high-energy x-ray diffraction measurements ($E$ = 99.54~keV; $\lambda$ = 0.01246~nm), using an area detector positioned 1597 mm behind the sample, were performed on the high energy station (6ID-D) in the MUCAT Sector at the Advanced Photon Source. At this high energy, x-rays probe the bulk of the sample rather than just the near-surface region and, by rocking the crystal about both the horizontal and vertical axes an extended range of a chosen reciprocal plane can be recorded\cite{kreyssig2007}. For these measurements, the samples were mounted on the cold-finger of a closed-cycle refrigerator using Kapton windows to avoid extraneous reflections associated with beryllium or the aluminum housing. For each data set, the sample was rotated over a range of horizontal angles, $\mu$, of $\pm$3.2~deg for each value of the vertical angle, $\eta$, between $\pm$3.2~deg with a step size of 0.4~deg. The splitting of the tetragonal $(220)_{\textrm{T}}$ diffraction peak into the $(400)_{\textrm{o}}/(040)_{\textrm{o}}$ orthorhombic diffraction peaks signals the tetragonal to orthorhombic distortion. The measured splitting was used to determine the dimensionless orthorhombicity parameter $\delta \equiv \left(  a_o - b_o \right)/\left( a_o + b_o \right)$, where $a_o$ and $b_o$ are the in-plane lattice constants of the orthorhombic unit cell.

Polarized-light imaging of domain structure was performed in a flow-type liquid $^4$He cryostat with the sample in vacuum. The sample was positioned on top of a polished copper cold finger and directly observed under linearly polarized light through an analyzer. The analyzer was turned, close but not exactly at 90$^o$ with respect to a polarizer. Due to the anisotropy of the optical properties in the orthorhombic phase, there is a rotation of the polarization plane upon reflection, always from $a-$ to $b-$ axis and the amount of the rotation scales with the orthorhombicity parameter $\delta$. Thus we observe a contrast between adjacent domains, because they rotate light in opposite directions.

The magnetization measurements were conducted in a \textit{Quantum Design} MPMS magnetometer. In the experiment, a platelet sample was fixed in a gelatine capsule with a small amount of Apiezon grease and the capsule was placed in a clear plastic straw. Care was taken to avoid artifact related to the tilt of the samples by repeated removal, re-mounting and re-measuring the same sample. At some concentrations several samples of different aspect ratio were measured. Then the measured magnetization was converted into the critical current density producing practically identical values indicating good quality of the samples and applicability of the Bean model. Critical current densities were determined from the measured magnetization following the Bean model \cite{Bean}, in which for a rectangular slab with the dimensions of $2c <2a \leq 2b$, $j_c\mathrm{ [A/cm^2]}=20M\mathrm{ [emu]}/\left[aV(1-a/(3b))\right]$ where all dimensions are in cm, $M$ is the total measured magnetic moment and $V$ is the sample volume. The obtained values are in a good agreement with those determined from the measurements of the magnetic induction profiles as well as from direct transport measurements as shown previously \cite{Prozorov}.  

The critical current density was compared at $T/T_c=0.7$ that was high enough to reach the field of the fishtail magnetization maximum, $H_{m}$, see Fig.~\ref{jcHT}, which has been used as a characteristic field. Another possibility was to evaluate the current in the remanent state, but the value of a magnetic moment at $H=$0 is significantly affected by the sample geometry due to demagnetization. Evaluation of the current density at $H_m$ has avoided such problem.
More details on the measurements and conversion procedures can be found elsewhere \cite{Prozorov}.

\section{Results}

\begin{figure}
\includegraphics[width=8.5cm]{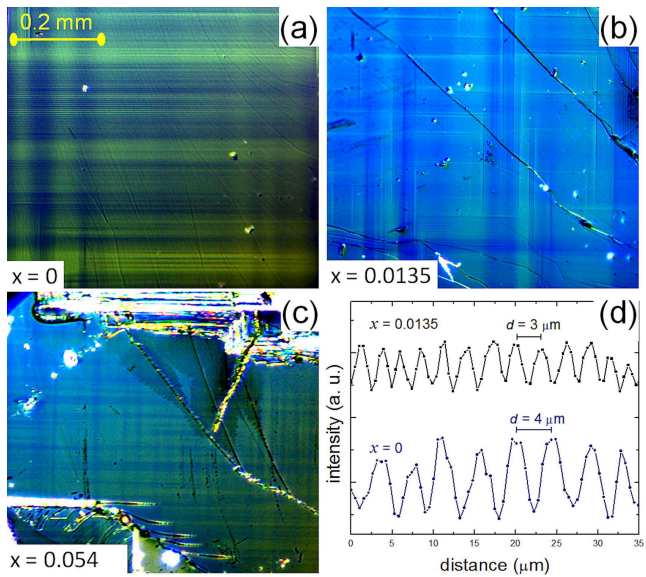}
\caption{ Polarized-light images showing structural domains at 5~K in Ba(Fe$_{1-x}$Co$_x$)$_2$As$_2$ crystals for $x$=0 (a), 0.0135 (b) and 0.054 (c). With increasing $x$, the domain wall separation $d$ becomes smaller (see intensity profiles across the boundaries in bottom right panel (d), giving $d \approx$4 $\mu$m for $x$=0 and $d \approx$3 $\mu$m for $x$=0.135. For $x \geq$0.024 we could not resolve individual domains since their width is below our digital camera resolution limit of $\sim$2 $\mu$m, however, the domains could be still tracked by eye. With further doping, the contrast of individual boundaries fades away and they become indistinguishable for $x$=0.047. However, wide stripes of domains ensembles can be seen for $x$ up to 0.054. Eventually the pattern becomes optically indistinguishable for $x \geq 0.058$ (not shown here). }
\label{domain_doping}
\end{figure}

In the previous study of domain formation in the parent compounds BaFe$_2$As$_2$, we showed that the domains are of 45$^o$ type, i.e. their walls are along the (110) direction in orthorhombic notation (or parallel to the tetragonal (100) axis) and that they span along the $c$-axis through a significant fraction of the crystal of 0.2 mm thickness. In Fig.~\ref{domain_doping} we show the evolution of the domain pattern with Co doping, $x$. As $x$ increases, several changes occur. (1) The characteristic separation between individual domain walls gets smaller, Fig.~\ref{domain_doping}(d), so we could not obtain a digital image that resolves individual boundaries already for $x \geq$ 0.024. However, we could still see them by eye up to $x<$ 0.047, with the resolution close to the optical limit $\sim 1 \mu$m set by the wavelength. On the other hand, the domains are large enough so that they do not cause X-ray Bragg diffraction spots to broaden above the instrumental resolution limit of about 150 nm. This gives an estimate of the characteristic width of the structural domains between 0.2 and 1 $\mu$m for $x$=0.047. (2) As a result, the density of the domain walls increases. (3) The same three typical domain patterns are found in parent and doped compounds, with vertical, horizontal and interwoven (crossing) boundaries. The interwoven patterns of domain bundles (not resolving the individual boundaries), shown in Fig.~\ref{domain_doping}(c) for $x$=0.054, become much more frequent with increasing $x$. (4) Optical contrast of domain boundaries fades away with an increase of $x$. 

\begin{figure}
\includegraphics[width=8.5cm]{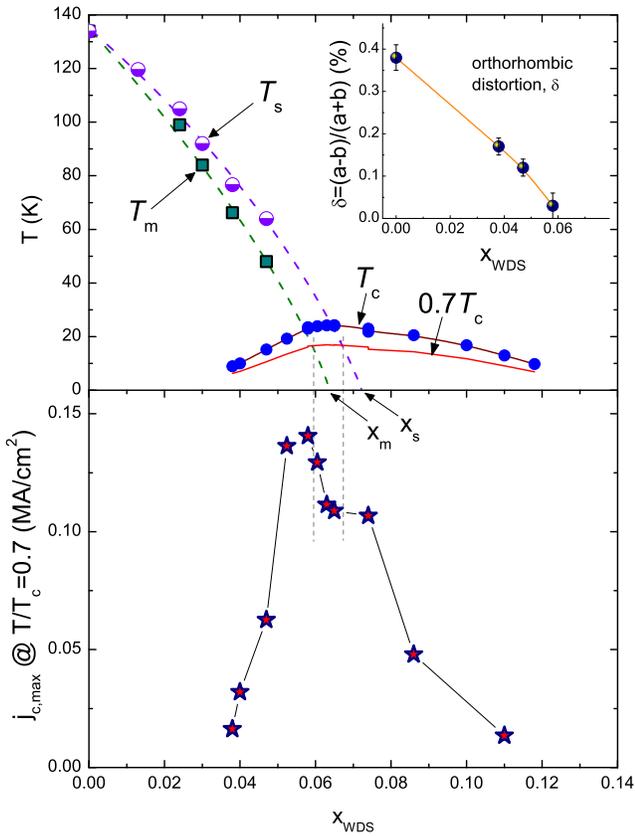}
\caption{Top panel. Doping phase diagram of the structural, $T_s$, the magnetic, $T_m$, transformations as well as  superconducting transition, $T_c$, in Ba(Fe$_{1-x}$Co$_x$)$_2$As$_2$ single crystals \cite{NiNiCo,pratt}. The $T_s(x)$ and $T_m(x)$ split with the increasing $x$. Red line inside the superconducting dome shows 0.7$T_c$ at which the critical current  (bottom panel) was evaluated. Inset in the top panel shows doping dependence of the degree of the orthorhombic distortion $\delta (x)$, taken at 7~K, base temperature of X-ray measurements. Bottom panel shows the evolution of the superconducting critical current density $j_{c,max}$ (stars). The critical current was evaluated at a fixed reduced temperature $T/T_c(x)$=0.7 (a dotted line in the top panel) and in the field corresponding to the maximum of the fishtail magnetization, see Fig.~\ref{jcHT}.
}
\label{phaseD}
\end{figure}

We could not optically resolve either individual domain boundaries or domain bundles in the compositions with $x>$0.054, despite the observations of anomalies in the temperature dependence of resistivity $\rho (T)$ accompanying structural transition at $T_s>T_c$ for sample with $x$=0.058 \cite{NiNiCo} and a notable suppression of the low temperature phonon part of thermal conductivity at temperatures below 1~K due to scattering off domain boundaries in samples with $x$ up to 0.074 \cite{thermalconductivity2}. The difficulty in resolving domains for $x$ close to optimal doping is primarily due to a suppression of the degree of the orthorhombic distortion, characterized by dimensionless parameter $\delta (x)$. The doping dependence of $\delta$ is shown in the inset in top panel of Fig.~\ref{phaseD}. The decreasing $\delta$ affects the optical contrast, since linearly-polarized light imaging detects the anisotropic optical response of the neighboring twins \cite{ZPhys1988} and the degree of this anisotropy is proportional to $\delta$. This decreases the difference in the light intensity between the twins. 

As can be seen from top panel of Fig.~\ref{phaseD} lines of magnetic, $T_m(x)$, and structural, $T_s(x)$, transitions coincide in the parent compound, $x=0$, and then separate with the increasing $x$ \cite{pratt,NiNiCo}. These lines can be extrapolated to $T=0$ to estimate the critical concentrations. Keeping in mind that the experimental data are confined to high temperatures with $T_s,~ T_m > T_c$ and thus the extrapolation is not precise, we estimate the critical concentration for the structural transition, $x_s \approx$0.070$\pm 0.03$, and for the magnetic transition, $x_m \approx$ 0.065$\pm 0.03$, see dashed lines in the top panel of Fig.~\ref{phaseD}. This opens a window of concentrations between $x_m$ and $x_s$ where only the structural distortion is present. The  X-ray measurements of the orthorhombic distortion indicate that $\delta (x)$, taken at a base temperature of 7~K, also decreases and extrapolates to zero in the same $x$ range within the experimental error, see the inset in Fig.~\ref{phaseD}. Importantly, since magnetic order in 122 pnictides has preferential direction in the ab-plane \cite{neutrons} and is rigidly correlated with the orthorhombic axes, structural domain walls are accompanied by the disruption of the magnetic order. In other words, magnetic domains follow the structural domains. A similar situation of correlated structural/magnetic  distortion in a magnetically ordered phase is found in the borocarbide superconductor ErNi$_2$B$_2$C \cite{Er-ortho,Detlefs}. Despite the uncertainty in the determination of the exact positions of $x_m$ and $x_s$, they are clearly quite close to the optimal concentration, so that superconductivity coexists with both types of order.

\begin{figure}
\includegraphics[width=8.5cm]{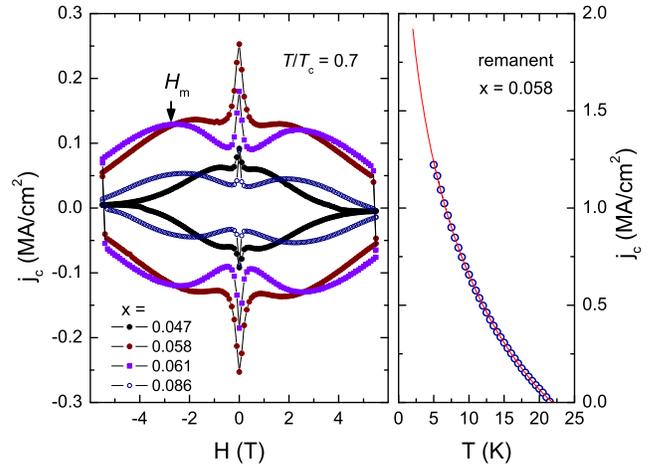}
\caption{Left panel. Evolution of the magnetization $M$ vs. $H$ loops with doping at $T/T_c=0.7$. The magnetization was converted into current density using Bean model \cite{Bean} to allow for a direct comparison of samples with different sizes. The data for different doping levels were taken at the same reduced temperature $T/T_c (x)$=0.7. Arrow shows the definition of $H_m$ (in this case for $x=$0.061) used to plot $j_{c,max}$ in Fig.~\ref{phaseD}. Right panel. The temperature dependence of the critical current density for sample with $x$=0.058, line shows a power-law fit.   }
\label{jcHT}
\end{figure}

In the bottom panel of Fig.~\ref{phaseD} we show the evolution of the superconducting critical current density $j_{c,max}$, taken for all doping levels at the same reduced temperature $T/T_c(x)$=0.7 at a magnetic field corresponding to the maximum in the $j_c(H)$ "fishtail" curve, $H_m$, as shown by the arrow for $x=$0.061 in Fig.~\ref{jcHT}. The magnetization curves are shown in the units of current density to allow for a direct comparison between samples of different sizes. Details of the conversion based on the Bean model \cite{Bean} for finite samples are given elsewhere \cite{Prozorov}. There is a clear asymmetry of $j_{c,max}(x)$ with respect to the $T_c (x)$ dome. As a function of doping, the critical current density starts to rise with decreasing $x$ below approximately $x_s$, peaks at $x_m$ or slightly below, and drops off quickly on moving towards $x=0$. 
The observed doping dependence of the critical current suggests that the best pinning conditions are realized in the presence of structural domains, and, perhaps, with the additional contribution of magnetic pinning. We conclude that domain walls act as effective pinning centers, and pinning becomes stronger for a finer and more interwoven domain structure and in the presence of static magnetism.

\section{Discussion}

As can be seen from Fig.~\ref{phaseD}, the critical current density as a function of doping has clear asymmetry and peaks in the underdoped regime. One may suggest that the observed enhancement is simply due to more robust superconductivity at the optimal doping. To refute this argument, let us consider two samples with similar $T_c$ on two sides of the dome. For example, for $x=0.054$, $T_c=$ 19.2 K and $j_c=1.36 \times 10^5$ A/cm$^2$, whereas in overdoped $x=0.086$ with even higher $T_c=$ 20.5 K, the critical current is substantially lower, $j_c=4.8 \times 10^4$ A/cm$^2$. Thus another reasons for the enhancement of $j_c$ should be looked for.

Comparison of top and bottom panels of Fig.~\ref{phaseD} suggests that the critical current peaks in the range where structural domains (1) become finely spaced, (2) their density increases, (3) they form interwoven patterns; (4) domains are more likely to be associated with both structural and magnetic order. As we will argue, ALL these features are favorable for pinning. First, the pinning strength is maximal when the size of the pinning centers (domain walls and their intersections) becomes comparable to the size of vortex cores. Second, volume pinning force increases with the increase of the density of the pinning centers. Domain pattern becomes finer with $x$ presumably due to lower energy cost of the domain wall formation with decreasing distortion $\delta$. Therefore, by moving towards finer and denser domain structure we both increase the pinning potential (more interwoven structures) and introduce more pinning centers per unit volume. Domains in Ba(Fe$_{1-x}$Co$_x$)$_2$As$_2$ propagate throughout the entire sample and, for $x$ close to optimal, form a dense maze of intersecting domain boundaries. A more frequent occurrence of interwoven patterns is also caused by the decreasing $\delta$. The areas of crossings are characterized by strong lattice deformation and their formation is energetically unfavorable as compared to parallel domain boundaries. Such natural linear defects extend along the c-axis, similar to heavy ion tracks. The nature is generous to at least 122 family of pnictide superconductors to create the defects of this type close to the optimal doping conditions where $T_c$ is the highest and the superconductivity is most robust.
The contribution of the magnetic order may also be very important for pinning, as suggested by the data in the bottom panel of Fig.~\ref{phaseD}. Because magnetic order in the planes requires reduced symmetry, orthorhombic distortion inevitably introduces disruptions of the long range magnetic order. Magnetic modulation itself creates strong modulation of the superconducting order parameter, acting as pinning centers. This modulation can be enhanced by the magnetostriction, as was shown by decoration imaging of vortex lattice in magnetic RNi$_2$B$_2$C \cite{Er-decoration,Er-ortho}.

The critical current density in Ba(Fe$_{1-x}$Co$_x$)$_2$As$_2$ reaches maximum at a composition with $x$=0.058. In addition, the critical current has a strong temperature dependence, as shown in the right panel of Fig.~\ref{jcHT}. An extrapolation to liquid helium temperatures gives values above the "technologically useful limit" of 1 MA/cm$^2$. This is almost 3 times higher than in the optimally doped crystals with $x$=0.074 \cite{Prozorov,equalpenetration}. Considering that the samples of Ba(Fe$_{1-x}$Co$_x$)$_2$As$_2$  in this range have upper critical fields as high as 60 T \cite{low_anisotropy,NiNiCo,Kano}, critical temperatures of above 20~K, low anisotropy \cite{low_anisotropy,singleton,Kano,tanatar_transport} and, as shown here, high intrinsic critical current densities, these materials can be considered as good candidates for applications. 

\section{Acknowledgements}

We thank C. Martin and V. G. Kogan for discussions. Work at the Ames Laboratory and at the MUCAT sector was supported by the Department of Energy - Basic Energy Sciences under Contract No. DE- AC02-07CH11358. Use of the Advanced Photon Source was supported by US DOE under Contract No. DE-AC02-06CH11357. R. P. acknowledges support from Alfred P. Sloan Foundation.


\end{document}